\documentclass[10pt,twocolumn,aps,prl,notitlepage,showpacs,floatfix,amsmath,amssymb;twocolumn]{revtex4-1}
\usepackage[latin9]{inputenc}
\usepackage{amsmath}
\usepackage{graphicx}
\usepackage{esint}

\makeatletter

\newcommand*\LyXThinSpace{\,\hspace{0pt}}

\@ifundefined{textcolor}{}
{%
 \definecolor{BLACK}{gray}{0}
 \definecolor{WHITE}{gray}{1}
 \definecolor{RED}{rgb}{1,0,0}
 \definecolor{GREEN}{rgb}{0,1,0}
 \definecolor{BLUE}{rgb}{0,0,1}
 \definecolor{CYAN}{cmyk}{1,0,0,0}
 \definecolor{MAGENTA}{cmyk}{0,1,0,0}
 \definecolor{YELLOW}{cmyk}{0,0,1,0}
}

\usepackage{float}

\usepackage{epstopdf}
 \usepackage{setspace}

\makeatother

\begin{document}

\title{Strongly correlated Fermi superfluid near an orbital Feshbach resonance:\\
 Stability, equation of state, and Leggett mode}

\author{Lianyi He$^{1}$, Jia Wang$^{2}$, Shi-Guo Peng$^{2,3}$, Xia-Ji
Liu$^{2}$, and Hui Hu$^{2}$}

\affiliation{$^{1}$State Key Laboratory of Low-Dimensional Quantum Physics and
Collaborative Innovation Center for Quantum Matter, Department of
Physics, Tsinghua University, Beijing 100084, China}

\affiliation{$^{2}$Centre for Quantum and Optical Science, Swinburne University
of Technology, Melbourne 3122, Australia}

\affiliation{$^{3}$State Key Laboratory of Magnetic Resonance and Atomic and
Molecular Physics, Wuhan Institute of Physics and Mathematics, Chinese
Academy of Science, Wuhan 430071, China}

\date{\today}
\begin{abstract}
We theoretically study the superfluid phase of a strongly correlated
$^{173}$Yb Fermi gas near its orbital Feshbach resonance, by developing
a quantitative pair-fluctuation theory within a two-band model. We
examine the density excitation spectrum of the system and determine
a stability phase diagram. We find that the $^{173}$Yb Fermi gas
is intrinsically metastable and has a peculiar equation of state, due
to the small but positive singlet scattering length. The massive Leggett
mode, arising from the fluctuation of the relative phase of two order
parameters, is severely damped. We discuss the parameter space where
an undamped Leggett mode may exist. 
\end{abstract}


\maketitle
%%%%%%%%%%%%%%%%%%%%%%%%%%%%%%%%%%%%%%%%%%%%%%%%%%%%%%%%%%%%%%%%%%%%%%%%%%%%%%%%%%%%%%%%%%%%%%%%

\section{Introduction}

\label{s1} %%%%%%%%%%%%%%%%%%%%%%%%%%%%%%%%%%%%%%%%%%%%%%%%%%%%%%%%%%%%%%%%%%%%%%%%%%%%%%%%%%%%%%%%%%%%%%%%

The realization of magnetic Feshbach resonance (MFR) in alkali-metal
atoms, i.e., tuning the $s$-wave scattering length of a two-component
atomic Fermi gas using a magnetic field \cite{Bloch2008,Chin2010},
opens a new paradigm for studying strongly correlated many-body phenomena.
The crossover from Bose-Einstein condensate (BEC) to Bardeen-Cooper-Schrieffer
(BCS) superfluid \cite{Giorgini2008} in both three \cite{Ketterle2008,Nascimbene2010,Horikoshi2010,Ku2012}
and two dimensions \cite{Martiyanov2010,Frohlich2011,Dyke2011,Murthy2015,Fenech2016}
has now been experimentally explored in greater detail, leading to
a number of new concepts such as a unitary fermionic superfluid and
universal equation of state (EoS) \cite{Ku2012,Ho2004,Hu2007} that
bring new insights to better understand other strongly interacting systems,
including high-$T_{c}$ superconductors \cite{Lee2006RMP}, nuclear
matter \cite{Lee2006PRA}, and quark-gluon plasma \cite{Kolb2004}.

For alkali-earth-metal atoms (such as Sr) or alkali-earth-metal-like atoms (i.e.,
Yb), however, the MFR mechanism does not work, due to their vanishing
total electron spin \cite{Chin2010}. In a recent pioneering work
by R. Zhang \textit{et al.} \cite{Zhang2015}, an alternative mechanism
of orbital Feshbach resonance (OFR) for $^{173}$Yb atoms has been
proposed. Because of a shallow bound state (i.e., a large triplet scattering
length) caused by the inter-orbital (nuclear) spin-exchange interactions,
the small difference in the nuclear Landé factor between different
orbital states allows the tunablity of scattering length through a
magnetic field \cite{Zhang2015}. The existence of the predicted OFR
has most recently been confirmed by either an anisotropic expansion
\cite{Pagano2015} or a cross-thermalization measurement \cite{Hofer2015},
which determined a resonance field $B_{0}=41\pm1$G \cite{Pagano2015}
or $B_{0}=55\pm8$G \cite{Hofer2015,Note}, respectively.

\begin{figure}
\begin{centering}
\includegraphics[width=0.48\textwidth]{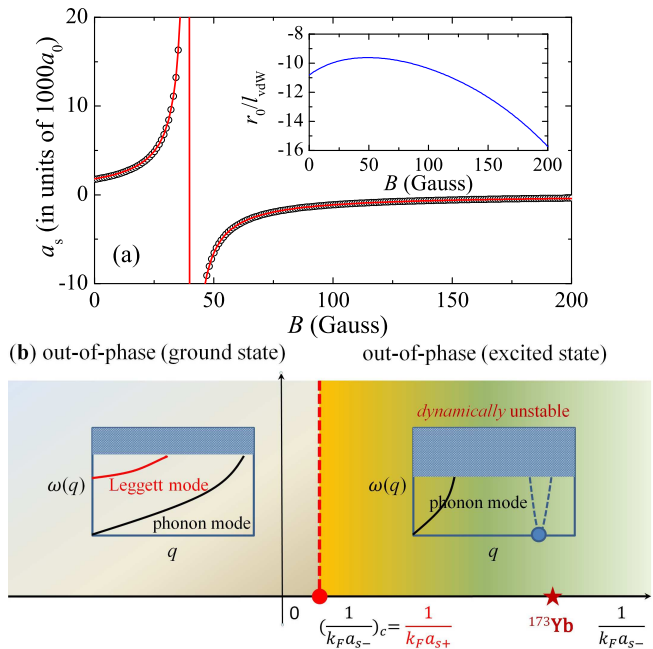} 
\par\end{centering}

\protect\protect\protect\caption{(color online). (a) The scattering length $a_{s}$ near the OFR of
$^{173}$Yb atoms. The circles are our two-body calculations, and the
red solid line is the fitting curve (see text). The inset shows the
effective range near the same resonance. (b) An illustration of the
many-body stability phase diagram. By tuning the interaction parameter
$1/(k_{{\rm F}}a_{s-})$ above a threshold $1/(k_{{\rm F}}a_{s+})$, where $a_{s-}$ and $a_{s+}$
are the singlet and triplet scattering lengths,
the out-of-phase solution, responsible for the OFR, develops an
anomalous mode in its low-energy (density) excitation spectrum and
is therefore dynamically unstable. In contrast, below the threshold,
the out-of-phase solution is stable and may host an undamped Leggett
mode.}

\label{fig1} 
\end{figure}

It is of great interest to explore the many-body physics of OFR. Indeed,
there are a number of urgent problems to address. Earlier qualitative
mean-field analysis introduced two order parameters and found that
the OFR is associated with the \emph{out-of-phase} solution of the
two pair potentials \cite{Zhang2015}. This solution is in fact an
excited state (saddle point) in the landscape of the thermodynamic
potential \cite{Zhai2016,Iskin2016}, and therefore may suffer from
the some instabilities encountered by the breached pairing or Sarma
phase in imbalanced Fermi gases \cite{Pao2006,He2009}. On the other
hand, the existence of two order parameters in OFR opens the possibility
of observing the long-sought \emph{massive} Leggett mode \cite{Leggett1966,Blumberg2007,Lin2012,Bittner2015}
resulted from the fluctuation of the relative phase of the two order
parameters. More fascinatingly, OFR is a narrow resonance due to the
significant closed-channel fraction \cite{Xu2016}. Would we observe
any peculiar feature of the EoS near the OFR of $^{173}$Yb atoms?

In this work, we address those interesting questions on stability,
equation of state and potential observation of the massive Leggett
mode, and present a \emph{quantitative} description of the zero-temperature
superfluid state of $^{173}$Yb atoms near OFR. Our main results are
briefly summarized as follows (see also Fig. \ref{fig1}). (1) Our two-body calculation with realistic
Lenard-Jones potentials predicts a resonance field $B_{0}\simeq39.4$
G [Fig. \ref{fig1}a], in good agreement with recent experimental observations \cite{Pagano2015,Hofer2015}.
(2) There is a dynamical instability revealed by the density excitation
spectrum [Fig. \ref{fig1}b]. Fortunately, due to the small singlet scattering length,
this instability occurs at very large momentum and hence is hard to
trigger under current experimental conditions. In other words, the
superfluid state of $^{173}$Yb atoms with OFR is intrinsically metastable.
(3) The small singlet scattering length also implies a peculiar EoS,
which is peculiar for a Feshbach resonance with sizable closed-channel
fraction. (4) The massive Leggett mode in a $^{173}$Yb Fermi gas
is severely damped. An undamped Leggett mode may exist only for the
case with both large singlet and triplet scattering lengths near the
OFR resonance [Fig. \ref{fig1}b].

%%%%%%%%%%%%%%%%%%%%%%%%%%%%%%%%%%%%%%%%%%%%%%%%%%%%%%%%%%%%%%%%%%%%%%%%%%%%%%%%%%%%%%%%%%%%%%%%

\section{Two-body calculation of $^{173}$Yb OFR}

\label{s2} %%%%%%%%%%%%%%%%%%%%%%%%%%%%%%%%%%%%%%%%%%%%%%%%%%%%%%%%%%%%%%%%%%%%%%%%%%%%%%%%%%%%%%%%%%%%%%%%

We start by briefly discussing the two-body physics for a Fermi gas
of $^{173}$Yb atoms with mass $M$ in different electronic (orbital)
states $^{1}S_{0}$ (denoted by $\left|g\sigma\right\rangle $) and
$^{3}P_{0}$ ($\left|e\sigma'\right\rangle $), where $\sigma$ and
$\sigma'$ stand for two nuclear spin states $\uparrow,\downarrow$.
In the absence of a magnetic field, a pair of atoms is well-described
using the single ($-$) or triplet ($+$) basis: 
\begin{equation}
\left|\pm\right\rangle =\frac{1}{2}(\left|ge\right\rangle \pm\left|eg\right\rangle )\otimes(\left|\uparrow\downarrow\right\rangle \mp\left|\downarrow\uparrow\right\rangle ).
\end{equation}
The interaction potentials are diagonal in this basis and are given
by Lenard-Jones potentials, 
\begin{equation}
V_{\pm}\left(r\right)=-\frac{C_{6}}{r^{6}}\left(1-\frac{\alpha_{\pm}^{6}}{r^{6}}\right),\label{eq:Lenard-Jones}
\end{equation}
where $C_{6}=2561$ a.u. for $^{173}$Yb \cite{Porsev2014} and $\alpha_{\pm}$
are the short-range parameters that are tuned to reproduce the singlet
scattering length $a_{s-}\simeq200a_{0}$ and the triplet scattering
length $a_{s+}\simeq1900a_{0}$ with $a_{0}$ being the Bohr radius
\cite{Hofer2015}. In the presence of magnetic field, due to the slightly
different Landé $g$ factor in two orbital states (i.e., $g_{g}\neq g_{e}$),
it is more convenient to introduce a two-channel description, with
the open- and closed-channel states given by 
\begin{eqnarray}
\left|{\rm o}\right\rangle  & = & \frac{1}{\sqrt{2}}(\left|-\right\rangle +\left|+\right\rangle ),\nonumber \\
\left|{\rm c}\right\rangle  & = & \frac{1}{\sqrt{2}}(\left|-\right\rangle -\left|+\right\rangle ).
\end{eqnarray}
One advantage of this new basis is that the Zeeman energy now becomes
diagonal, and their difference in the two channels is $\delta(B)=\delta\mu B$,
where $\delta\mu=(g_{e}-g_{g})(m_{\uparrow}-m_{\downarrow})\mu_{B}=2\pi\hbar\times112\Delta_{m}$
Hz/G with the Bohr magneton $\mu_{B}$ and $\Delta_{m}=5$ \cite{Pagano2015,Hofer2015}.
The key advantage, however, is the brilliant idea \cite{Zhang2015}
that the scattering length in the open channel could be tuned by varying
the detuning $\delta(B)$, exactly analogous to a MFR, provided that
the bound-state energy in the closed channel is comparable to $\delta(B)$.
This condition is generally impossible to satisfy, since $\delta(B)$
for nuclear spins is typically several order smaller in magnitude
than that in a MFR. Luckily, for $^{173}$Yb atoms, the \emph{shallow}
bound state due to the large triplet scattering length $a_{s+}$ has
the desired energy scale $\sim\delta(B)$.

The existence of such an OFR has been theoretically examined by using
the pseudo-potential approach and the finite-range potential model
\cite{Zhang2015}. In Ref. \cite{Hofer2015}, by using a low-energy
expansion of the singlet and triplet scattering phase shifts, where
the effective ranges based on realistic potentials were included,
the resonance field was predicted to be $B_{0}\simeq42$G. Here, we
present a more realistic calculation by using the Lenard-Jones potential
Eq. (\ref{eq:Lenard-Jones}) and standard $R$-matrix propagation
method \cite{Baluja1982}, as shown in Fig. 1(a). We find that the scattering
observable in the open channel such as the scattering length $a_{s}$
is not sensitive to $\alpha_{\pm}$ as long as $a_{s\pm}$ are reproduced.
The calculated scattering length in the open channel is well fitted
by a simple expression, 
\begin{equation}
a_{s}=a_{\textrm{bg}}-\frac{\bar{a}\bar{E}s_{\textrm{res}}}{\delta\mu\left(B-B_{0}\right)},
\end{equation}
with the parameters 
\begin{equation}
a_{\textrm{bg}}\simeq29.96a_{0},\ \ \ \ \ \ s_{\textrm{res}}\simeq0.154.
\end{equation}
The resonance field $B_{0}$ is predicted to be 
\begin{equation}
B_{0}\simeq39.4{\rm G}.
\end{equation}
Here, $\bar{a}\equiv[4\pi/\Gamma(1/4)^{2}]l_{\textrm{vdW}}$ and $\bar{E}=1/(M\bar{a}^{2})$
is the length and energy related to the van-der-Waals length $l_{\textrm{vdW}}\equiv(1/2)(MC_{6})^{1/4}\simeq84.8a_{0}$
and we set $\hbar=1$. We find that the predicted resonance field
$B_{0}\simeq39.4$ G agrees well with the experimental measurements
\cite{Pagano2015,Hofer2015}. We note that, the small $s_{\textrm{res}}$
implies that the OFR of $^{173}$Yb atoms is a closed-channel dominated
scattering \cite{Chin2010}.

%%%%%%%%%%%%%%%%%%%%%%%%%%%%%%%%%%%%%%%%%%%%%%%%%%%%%%%%%%%%%%%%%%%%%%%%%%%%%%%%%%%%%%%%%%%%%%%%

\section{Effective field theory of OFR}

\label{s3} %%%%%%%%%%%%%%%%%%%%%%%%%%%%%%%%%%%%%%%%%%%%%%%%%%%%%%%%%%%%%%%%%%%%%%%%%%%%%%%%%%%%%%%%%%%%%%%%

The minimal model Hamiltonian for OFR can be given by $\mathcal{H}=\mathcal{H}_{0}+\mathcal{H}_{I}$,
where 
\begin{eqnarray}
\mathcal{H}_{0} & = & \sum_{{\rm n}i}\int d\mathbf{r}\psi_{{\rm n}i}^{\dagger}\left(\mathbf{r}\right)\left(-\frac{\nabla^{2}}{2M}+\varepsilon_{{\rm n}i}\right)\psi_{{\rm n}i}\left(\mathbf{r}\right),\label{two-band}\\
\mathcal{H}_{I} & = & \sum_{{\rm nm}}\int d\mathbf{r}d\mathbf{r}'\varphi_{{\rm n}}^{\dagger}\left(\mathbf{r}\right)V_{{\rm nm}}\left(\left|\mathbf{r}-\mathbf{r}'\right|\right)\varphi_{{\rm m}}\left(\mathbf{r'}\right).
\end{eqnarray}
Here $\varphi_{{\rm n}}(\mathbf{r})=\psi_{{\rm n}2}(\mathbf{r})\psi_{{\rm n}1}\left(\mathbf{r}\right)$,
and the subscript ${\rm n}={\rm o,c}$ denotes the open or closed
channel. The two internal degrees of freedom in each channel are indicated
by $i=1,2$. Without loss of generality, the threshold energies $\varepsilon_{{\rm n}i}$
can be chosen as 
\begin{eqnarray}
\varepsilon_{{\rm o}1}=\varepsilon_{{\rm o}2}=0,\ \ \ \ \ \ \varepsilon_{{\rm c}1}=\varepsilon_{{\rm c}2}=\frac{1}{2}\delta(B).
\end{eqnarray}
The interaction potentials $V_{{\rm nm}}(r)$ following the basis
transformation of Eq. (\ref{eq:Lenard-Jones}) read, 
\begin{eqnarray}
V_{{\rm oo}}(r)=V_{{\rm cc}}(r) & = & \frac{1}{2}[V_{-}(r)+V_{+}(r)],\nonumber \\
V_{{\rm oc}}(r)=V_{{\rm co}}(r) & = & \frac{1}{2}[V_{-}(r)-V_{+}(r)].
\end{eqnarray}

The realistic form of the microscopic potential $V_{{\rm nm}}(r)$
is rather hard for both the scattering problem and the many-body problem.
The effective ranges $r_{\pm}$ of the microscopic potentials $V_{\pm}(r)$
introduces an energy scale $\varepsilon_{{\rm r}}\sim1/(Mr_{\pm}^{2})$.
At low scattering energy $E=k^{2}/M\ll\varepsilon_{{\rm r}}$, the
shape of the microscopic interaction potentials $V_{\pm}(r)$ is not
important. For many-body physics, this means that all kinds of short-ranged
potentials $V_{\pm}(r)$ with the same scattering lengths $a_{s\pm}$
lead to the same prediction in the dilute limit. One way to simplify
the calculation is to use the pseudo-potentials \cite{Zhang2015}
\begin{eqnarray}
V_{\pm}(r)\simeq\frac{4\pi a_{s\pm}}{M}\delta({\bf r})\frac{\partial}{\partial r}(r\cdot),
\end{eqnarray}
or equivalently 
\begin{eqnarray}
V_{{\rm oo}}(r) & = & V_{{\rm cc}}(r)\simeq\frac{4\pi a_{s0}}{M}\delta({\bf r})\frac{\partial}{\partial r}(r\cdot),\nonumber \\
V_{{\rm oc}}(r) & = & V_{{\rm co}}(r)\simeq\frac{4\pi a_{s1}}{M}\delta({\bf r})\frac{\partial}{\partial r}(r\cdot).
\end{eqnarray}
Here the scattering lengths $a_{s0}$ and $a_{s1}$ are defined as
\begin{eqnarray}
a_{s0}=\frac{1}{2}(a_{s-}+a_{s+}),\ \ \ \ \ a_{s1}=\frac{1}{2}(a_{s-}-a_{s+}).
\end{eqnarray}

However, for the purpose of making use of the field theoretical approaches
for the many-body problem, it is more convenient to employ the leading-order
low-energy effective theory, i.e., the contact interaction potential.
Therefore, we write 
\begin{equation}
V_{{\rm nm}}\left(\left|\mathbf{r}-\mathbf{r}'\right|\right)=V_{{\rm nm}}\delta\left(\mathbf{r}-\mathbf{r}'\right).
\end{equation}
Here the contact couplings $V_{{\rm oo}}=V_{{\rm cc}}$ and $V_{{\rm oc}}=V_{{\rm co}}$
are bare quantities and should be renormalized by using the physical
scattering lengths $a_{s\pm}$ or $a_{s0,1}$. By making use of the
contact potentials, the Lippmann-Schwinger equation of the scattering
$T$ matrix becomes a simple algebra equation, 
\begin{eqnarray}
 &  & \left(\begin{array}{cc}
T_{{\rm oo}}(E) & T_{{\rm oc}}(E)\\
T_{{\rm co}}(E) & T_{{\rm cc}}(E)
\end{array}\right)^{-1}\nonumber \\
 & = & \left(\begin{array}{cc}
V_{{\rm oo}} & V_{{\rm oc}}\\
V_{{\rm co}} & V_{{\rm cc}}
\end{array}\right)^{-1}-\left(\begin{array}{cc}
{\cal B}_{{\rm o}}(E) & 0\\
0 & {\cal B}_{{\rm c}}(E)
\end{array}\right),\label{Lippmann}
\end{eqnarray}
where the two-particle bubble functions are given by 
\begin{eqnarray}
{\cal B}_{{\rm o}}(E) & = & \sum_{{\bf p}}\frac{1}{E+i\epsilon-2\varepsilon_{{\bf p}}},\nonumber \\
{\cal B}_{{\rm c}}(E) & = & \sum_{{\bf p}}\frac{1}{E+i\epsilon-\delta(B)-2\varepsilon_{{\bf p}}}
\end{eqnarray}
Here $\epsilon=0^{+}$ and $\varepsilon_{{\bf p}}={\bf p}^{2}/(2M)$.
The cost of the contact interaction is that the integral over the
fermion momentum ${\bf p}$ becomes divergent. We introduce a cutoff
$\Lambda$ for $|{\bf p}|$ and obtain 
\begin{eqnarray}
{\cal B}_{{\rm o}}(E) & = & -\eta(\Lambda)+\Pi_{{\rm o}}(E),\nonumber \\
{\cal B}_{{\rm c}}(E) & = & -\eta(\Lambda)+\Pi_{{\rm c}}(E),
\end{eqnarray}
where the divergent pieces read 
\begin{equation}
\eta(\Lambda)=\sum_{{\bf p}}\frac{1}{2\varepsilon_{{\bf p}}}=\frac{M\Lambda}{2\pi^{2}}.
\end{equation}
The finite pieces are given by 
\begin{eqnarray}
\Pi_{{\rm o}}(E) & = & \frac{M}{4\pi}\sqrt{-M(E+i\epsilon)},\nonumber \\
\Pi_{{\rm c}}(E) & = & \frac{M}{4\pi}\sqrt{-M(E+i\epsilon-\delta)}.
\end{eqnarray}
Physically, the UV cutoff $\Lambda$ corresponds to the momentum scale
of order of $O(1/r_{\pm})$ and should be sent to infinity if we set
$r_{\pm}\rightarrow0$.

%%%%%%%%%%%%%%%%%%%%%%%%%%%%%%%%%%%%%%%%%%%%%%%%%%%%%%%%%%%%%%%%%%%%%%%%%%%%%%%%%%%%%%%%%%%%%%%%

\subsection{Renormalization}

%%%%%%%%%%%%%%%%%%%%%%%%%%%%%%%%%%%%%%%%%%%%%%%%%%%%%%%%%%%%%%%%%%%%%%%%%%%%%%%%%%%%%%%%%%%%%%%%

The UV divergence can be completely removed by renormalization of
the bare contact coupling matrix $V$. The renormalized coupling matrix
$U$ is related to the bare coupling matrix through \cite{He2015}
\begin{eqnarray}
\left(\begin{array}{cc}
U_{{\rm oo}} & U_{{\rm oc}}\\
U_{{\rm co}} & U_{{\rm cc}}
\end{array}\right)^{-1}=\left(\begin{array}{cc}
V_{{\rm oo}} & V_{{\rm oc}}\\
V_{{\rm co}} & V_{{\rm cc}}
\end{array}\right)^{-1}+\eta(\Lambda)I_{2\times2}.\label{Lippmann}
\end{eqnarray}
Therefore, we have 
\begin{equation}
U_{{\rm oo}}=U_{{\rm cc}}\equiv U_{0},\ \ \ \ U_{{\rm oc}}=U_{{\rm co}}\equiv U_{1}.
\end{equation}
Then the Lippmann-Schwinger equation becomes cutoff independent:
\begin{eqnarray}
 &  & \left(\begin{array}{cc}
T_{{\rm oo}}(E) & T_{{\rm oc}}(E)\\
T_{{\rm co}}(E) & T_{{\rm cc}}(E)
\end{array}\right)^{-1}\nonumber \\
 & = & \left(\begin{array}{cc}
U_{0} & U_{1}\\
U_{1} & U_{0}
\end{array}\right)^{-1}-\left(\begin{array}{cc}
\Pi_{{\rm o}}(E) & 0\\
0 & \Pi_{{\rm c}}(E)
\end{array}\right).
\end{eqnarray}
Solving the Lippmann-Schwinger equation, we obtain the $T$ matrix
for the open channel, 
\begin{eqnarray}
T_{{\rm oo}}^{-1}(E)=\left[U_{0}+\frac{U_{1}^{2}\Pi_{{\rm c}}(E)}{1-U_{0}\Pi_{{\rm c}}(E)}\right]^{-1}-\Pi_{{\rm o}}(E).
\end{eqnarray}

To complete the contact potential description of the orbital Feshbach
resonance, we finally need to relate the elements of the renormalized
coupling matrix $U$ to the physical quantities. To this end, we calculate the
open-channel scattering amplitude 
\begin{eqnarray}
f_{{\rm o}}(k)=-\frac{M}{4\pi}T_{{\rm oo}}\left(E=\frac{k^{2}}{M}\right).
\end{eqnarray}
It can be expressed as 
\begin{eqnarray}
f_{{\rm o}}(k)=\frac{1}{k\cot\delta_{s}(k)-ik},
\end{eqnarray}
where the effective $s$-wave scattering phase shift $\delta_{s}(k)$
is given by 
\begin{equation}
k\cot\delta_{s}(k)=-\frac{1-\frac{MU_{0}}{4\pi}\sqrt{M\delta-k^{2}}}{\frac{MU_{0}}{4\pi}-\left[\left(\frac{MU_{0}}{4\pi}\right)^{2}-\left(\frac{MU_{1}}{4\pi}\right)^{2}\right]\sqrt{M\delta-k^{2}}}
\end{equation}
Matching this result to the known result from quantum mechanical calculation
\cite{Zhang2015}, we obtain 
\begin{equation}
U_{0}=\frac{4\pi a_{s0}}{M},\ \ \ \ U_{1}=\frac{4\pi a_{s1}}{M}.
\end{equation}
The effective $s$-wave scattering length of the open channel can
be given by $a_{s}=-f_{{\rm o}}(k=0)$. We obtain \cite{Zhang2015}
\begin{equation}
a_{s}=\frac{a_{s0}-(a_{s0}^{2}-a_{s1}^{2})\sqrt{M\delta}}{1-a_{s0}\sqrt{M\delta}}.
\end{equation}
Therefore, there exists a scattering resonance at $\delta=1/(Ma_{s0}^{2})$
if $a_{s0}>0$ \cite{Zhang2015}.

%%%%%%%%%%%%%%%%%%%%%%%%%%%%%%%%%%%%%%%%%%%%%%%%%%%%%%%%%%%%%%%%%%%%%%%%%%%%%%%%%%%%%%%%%%%%%%%%

\subsection{Bound states}

%%%%%%%%%%%%%%%%%%%%%%%%%%%%%%%%%%%%%%%%%%%%%%%%%%%%%%%%%%%%%%%%%%%%%%%%%%%%%%%%%%%%%%%%%%%%%%%%

The bound states or molecule states can be obtained by solving the
poles of the off-shell $T$ matrix $T(Z)$ with the on-shell scattering
energy $E$ replaced with the off-shell variable $Z=\omega-{\bf q}^{2}/(4M)$.
Here $\omega$ and ${\bf q}$ represents the energy and momentum of
the two-body states, respectively. The bound states corresponds to
the $Z<0$ poles of the following equation:
\begin{equation}
\det\left(\begin{array}{cc}
T_{{\rm oo}}(Z) & T_{{\rm oc}}(Z)\\
T_{{\rm co}}(Z) & T_{{\rm cc}}(Z)
\end{array}\right)^{-1}=0,
\end{equation}
or explicitly, 
\begin{eqnarray}
 &  & \frac{1}{a_{{\rm s}0}^{2}-a_{{\rm s}1}^{2}}-\frac{a_{{\rm s}0}}{a_{{\rm s}0}^{2}-a_{{\rm s}1}^{2}}\left[\sqrt{-MZ}+\sqrt{-M(Z-\delta)}\right]\nonumber \\
 &  & +\sqrt{-MZ}\sqrt{-M(Z-\delta)}=0.
\end{eqnarray}

Since the OFR exists only if $a_{s0}>0$, we set $a_{s0}>0$, and hence
the resonance point is $\delta_{{\rm res}}=1/(Ma_{{\rm s}0}^{2})$.
By making use of $\delta_{{\rm res}}$, we can express the pole equation
as 
\begin{eqnarray}
\frac{1-\sqrt{-x}-\sqrt{-x+d}}{1-t^{2}}+\sqrt{-x(-x+d)}=0.\label{bound-state}
\end{eqnarray}
Here the dimensionless variables are defined as $x=Z/\delta_{{\rm res}}$
and $d=\delta/\delta_{{\rm res}}$. It is clear that the energy spectrum
of the bound states depends solely on the ratio 
\begin{eqnarray}
t=\frac{a_{s1}}{a_{s0}}=\frac{a_{s-}-a_{s+}}{a_{s-}+a_{s+}}.
\end{eqnarray}
Since $a_{s0}>0$ we can set $a_{s+}>0$ without loss of generality.
We have $a_{s-}/a_{s+}>-1$ and therefore $-\infty<t<1$. We therefore
find two cases for the bound state spectrum: \\
(1) If $a_{s-}>0$ and hence $-1<t<1$ or $t^{2}<1$, there exist
two molecule states: One is the Feshbach molecule state, which exists
at the BEC side of the resonance $0<\delta<\delta_{{\rm res}}$, and
the other is a bound state below the Feshbach molecule state, which
exists for all values of $\delta$. A special case is $t=0$ which
means the two channels decouples. We have two solutions: $Z=-\delta_{{\rm res}}$
which exists for all $\delta$, and $Z=\delta-\delta_{{\rm res}}$
which exists for $0<\delta<\delta_{{\rm res}}$. \\
 (2) If $a_{s-}<0$ and hence $t<-1$ or $t^{2}>1$, the pole equation
gives only one solution at the BEC side $0<\delta<\delta_{{\rm res}}$,
corresponding the Feshbach molecule state.

To understand the above results (and also for the understanding of
the many-body case), it is intuitive to take a look at the case $\delta=0$.
In this case, Eq. (\ref{bound-state}) can be simplified as 
\begin{equation}
(1\pm|t|)\sqrt{-x}=1.
\end{equation}
Therefore, for $|t|<1$ or $a_{s-}>0$, there exist two solutions
\begin{equation}
Z_{\pm}(0)=-\left(\frac{1}{1\pm|t|}\right)^{2}\delta_{{\rm res}}.
\end{equation}
For $|t|\rightarrow1$, we have $|Z_{-}(0)|\gg|Z_{+}(0)|$, and hence
the two bound state levels are well separated. In this case, the solution
$Z_{-}$, which is almost a constant for all values of the detuning
$\delta$, corresponds to a deep bound state and may decouple from
the BCS-BEC crossover physics. For $^{173}$Yb atoms, we have $a_{s+}\simeq1900a_{0}$
and $a_{s-}\simeq200a_{0}$ and hence $t\simeq-0.81$. In this case,
the two solutions are given by 
\begin{equation}
Z_{\pm}(0)=-\frac{1}{Ma_{s\pm}^{2}}.\label{Binding0}
\end{equation}
Therefore, for $^{173}$Yb atoms we have 
\begin{equation}
\frac{|Z_{-}(0)|}{|Z_{+}(0)|}\simeq90.
\end{equation}
A full energy spectrum in the range $0<\delta/\delta_{{\rm res}}<2$
is shown in Fig. 2. It is clear that a Feshbach molecule state ($Z_{+}$)
exists in the BEC regime ($0<\delta<\delta_{{\rm res}}$). Another
deep bound state $Z_{-}$ exists for all values of $\delta$.

\begin{figure}
\begin{centering}
\includegraphics[width=0.48\textwidth]{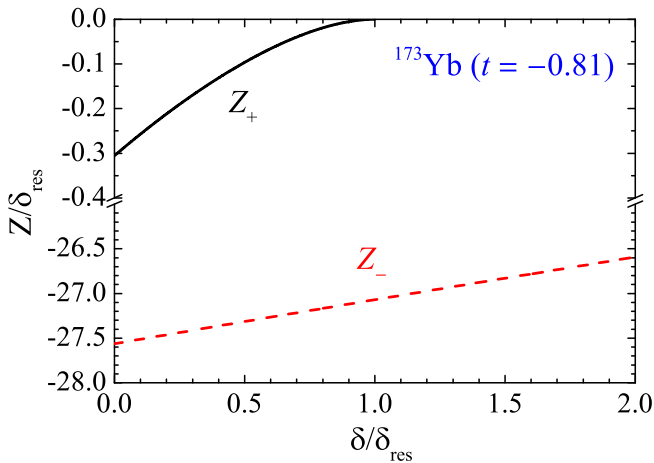} 
\par\end{centering}

\protect\protect\protect\caption{(color online). Energy spectrum of the bound states across the OFR
in $^{137}$Yb atoms. The two energy levels, the Feshbach molecule
state ($Z_{+}$) and the deep bound state ($Z_{-}$), are determined
by Eq. (\ref{bound-state}).}

\label{fig2} 
\end{figure}

%%%%%%%%%%%%%%%%%%%%%%%%%%%%%%%%%%%%%%%%%%%%%%%%%%%%%%%%%%%%%%%%%%%%%%%%%%%%%%%%%%%%%%%%%%%%%%%%
\section{Many-body theory of Fermi gases across an OFR}
\label{s4} 
%%%%%%%%%%%%%%%%%%%%%%%%%%%%%%%%%%%%%%%%%%%%%%%%%%%%%%%%%%%%%%%%%%%%%%%%%%%%%%%%%%%%%%%%%%%%%%%%

The two-band model (\ref{two-band}) that uses the singlet and triplet
scattering lengths $a_{s\pm}$ as the input provides a \emph{minimal}
model to describe the many-body aspect of OFR \cite{Iskin2016,He2015}.
In the dilute limit, it agrees reasonably with the two-body calculation
\cite{Zhang2015,Xu2016}, and within the mean-field approximation
it captures the qualitative physics of superfluid pairings \cite{Zhang2015,Iskin2016}.
Here, we consider \emph{strong} pair fluctuations on top of the mean-field
solution, which must be accounted for near OFR. The grand canonical
Hamiltonian of the two-band model is given by 
\begin{eqnarray}
\mathcal{H}-\mu\mathcal{N} & = & \sum_{{\rm n}i}\int d\mathbf{r}\psi_{{\rm n}i}^{\dagger}\left(\mathbf{r}\right)\left(-\frac{\nabla^{2}}{2M}-\mu_{{\rm n}}\right)\psi_{{\rm n}i}\left(\mathbf{r}\right)\nonumber \\
 &  & +\sum_{{\rm nm}}V_{{\rm nm}}\int d\mathbf{r}\varphi_{{\rm n}}^{\dagger}\left(\mathbf{r}\right)\varphi_{{\rm m}}\left(\mathbf{r}\right).
\end{eqnarray}
Here $\mu$ is the chemical potential conjugated to the total particle
number $\mathcal{N}=\sum_{{\rm n}i}\int d\mathbf{r}\psi_{{\rm n}i}^{\dagger}\left(\mathbf{r}\right)\psi_{{\rm n}i}\left(\mathbf{r}\right)$.
The effective chemical potentials of the two channels are defined
as 
\begin{equation}
\mu_{{\rm o}}=\mu,\ \ \ \ \ \ \mu_{{\rm c}}=\mu-\frac{1}{2}\delta(B).
\end{equation}

We solve the two-band model Hamiltonian by using a functional path-integral
approach \cite{He2015,SadeMelo1993,Hu2006,Diener2008,He2015-2,Bighin2016}.
The partition function of the many-body system can be expressed as
\begin{eqnarray}
{\cal Z}=\int[d\psi][d\psi^{\dagger}]\exp{\left(-{\cal S}\right)},
\end{eqnarray}
where the action ${\cal S}$ reads 
\begin{eqnarray}
{\cal S}=\int dx\sum_{{\rm n}i}\psi_{{\rm n}i}^{\dagger}(x)\partial_{\tau}\psi_{{\rm n}i}^{\phantom{\dag}}(x)+\int_{0}^{\beta}d\tau(\mathcal{H}-\mu\mathcal{N}).
\end{eqnarray}
Here $x=(\tau,{\bf r})$ and $\int dx=\int_{0}^{\beta}d\tau\int d^{3}{\bf r}$,
with $\tau$ being the imaginary time, and $\beta=1/T$, with $T$
being the temperature of the system and the Boltzmann constant $k_{{\rm B}}=1$.
Following the standard field theoretical treatment, we introduce the
auxiliary pairing fields 
\begin{eqnarray}
\Phi(x)=\left(\begin{array}{c}
\Phi_{{\rm o}}(x)\\
\Phi_{{\rm c}}(x)
\end{array}\right)=\left(\begin{array}{cc}
V_{{\rm oo}} & V_{{\rm oc}}\\
V_{{\rm co}} & V_{{\rm cc}}
\end{array}\right)\left(\begin{array}{c}
\varphi_{{\rm o}}(x)\\
\varphi_{{\rm c}}(x)
\end{array}\right),
\end{eqnarray}
apply the Hubbard-Stratonovich transformation, and integrate out the
fermion fields. The partition function of the system can be expressed
as 
\begin{equation}
{\cal Z}=\int[d\Phi][d\Phi^{\dagger}]\exp{\left(-{\cal S}_{{\rm eff}}\right)}.
\end{equation}
The effective action ${\cal S}_{{\rm eff}}$ reads 
\begin{eqnarray}
{\cal S}_{{\rm eff}} & = & -\int dx\ \Phi^{\dagger}(x)V^{-1}\Phi(x)\nonumber \\
 &  & -\sum_{{\rm n}={\rm o,c}}{\rm Tr}\ln{\bf G}_{{\rm n}}^{-1}[\Phi_{{\rm n}}(x)],
\end{eqnarray}
where the inverse fermion Green's functions are given by 
\begin{eqnarray}
{\bf G}_{{\rm n}}^{-1} & = & \left(\begin{array}{cc}
-\partial_{\tau}+\frac{\nabla^{2}}{2M}+\mu_{{\rm n}} & \Phi_{{\rm n}}(x)\\
\Phi_{{\rm n}}^{*}(x) & -\partial_{\tau}-\frac{\nabla^{2}}{2M}-\mu_{{\rm n}}
\end{array}\right)\nonumber \\
 &  & \times\delta(x-x^{\prime}).
\end{eqnarray}

In the superfluid phase, the pairing fields have nonzero expectation
values. We write 
\begin{equation}
\Phi_{{\rm n}}(x)=\Delta_{{\rm n}}+\phi_{{\rm n}}(x),
\end{equation}
where the uniform parts $\Delta_{{\rm o}}$ and $\Delta_{{\rm c}}$
serve as the order parameters of superfluidity. The effective action
${\cal S}_{{\rm eff}}$ can then be expanded about its mean-field
solution, or in powers of the quantum fluctuations $\phi_{{\rm o}}(x)$
and $\phi_{{\rm c}}(x)$, leading to \cite{He2015,SadeMelo1993,Hu2006,Diener2008}
\begin{equation}
{\cal S}_{{\rm eff}}\left[\Phi,\Phi^{*}\right]={\cal S}_{{\rm MF}}+{\cal S}_{{\rm GF}}\left[\phi,\phi^{*}\right]+\dots.
\end{equation}
Here ${\cal S}_{{\rm MF}}$ is the mean-field part, and ${\cal S}_{{\rm GF}}\left[\phi,\phi^{*}\right]$
denotes the Gaussian fluctuation part, which is quadratic in $\phi$
and $\phi^{*}$. In the Gaussian pair fluctuation (GPF) theory, all
the fluctuation contributions beyond Gaussian are neglected.

The mean-field contribution to the thermodynamic potential, $\Omega_{{\rm MF}}={\cal S}_{{\rm MF}}/(\beta V)$,
is given by 
\begin{equation}
\Omega_{\textrm{MF}}=-\mathbf{\Delta}^{\dagger}\left(\begin{array}{cc}
V_{{\rm oo}} & V_{{\rm oc}}\\
V_{{\rm co}} & V_{{\rm cc}}
\end{array}\right)^{-1}\mathbf{\Delta}+\sum_{{\rm n}\mathbf{k}}\left(\xi_{{\rm n}\mathbf{k}}-E_{{\rm n}\mathbf{k}}\right),
\end{equation}
where $\mathbf{\Delta}\equiv(\Delta_{{\rm o}},\Delta_{{\rm c}})^{T}$
and the dispersions in each channel are defined as $\xi_{{\rm n}\mathbf{k}}=\varepsilon_{\mathbf{k}}-\mu_{{\rm n}}$
and $E_{{\rm n}\mathbf{k}}=\sqrt{\xi_{{\rm n}\mathbf{k}}^{2}+|\Delta_{{\rm n}}|^{2}}$.
By using the renormalized coupling matrix $U$, we find that the UV
divergence is completely removed. We obtain 
\begin{equation}
\Omega_{\textrm{MF}}=-\mathbf{\Delta}^{\dagger}\left(\begin{array}{cc}
\lambda_{0} & \lambda_{1}\\
\lambda_{1} & \lambda_{0}
\end{array}\right)\mathbf{\Delta}+\sum_{{\rm n}\mathbf{k}}\left(\xi_{{\rm n}\mathbf{k}}-E_{{\rm n}\mathbf{k}}+\frac{\Delta_{{\rm n}}^{2}}{2\varepsilon_{\mathbf{k}}}\right),\label{eq:OmegaMF}
\end{equation}
where 
\begin{equation}
\lambda_{0}=\frac{M}{4\pi}\frac{a_{s0}}{a_{s0}^{2}-a_{s1}^{2}},\ \ \ \ \lambda_{1}=-\frac{M}{4\pi}\frac{a_{s1}}{a_{s0}^{2}-a_{s1}^{2}}.
\end{equation}
In the GPF theory, the order parameters $\Delta_{{\rm o}}$ and $\Delta_{{\rm c}}$
as functions of the chemical potential $\mu$ should be determined
by the stationary condition $\partial\Omega_{\textrm{MF}}/\partial\Delta_{{\rm n}}=0$,
which gives rise to the so-called gap equation, 
\begin{equation}
\left[\begin{array}{cc}
F_{{\rm o}}\left(\Delta_{{\rm o}}\right) & -\lambda_{1}\\
-\lambda_{1} & F_{{\rm c}}\left(\Delta_{{\rm c}}\right)
\end{array}\right]\left(\begin{array}{c}
\Delta_{{\rm o}}\\
\Delta_{{\rm c}}
\end{array}\right)=0,\label{gapEq}
\end{equation}
where 
\begin{equation}
F_{{\rm n}}(\Delta_{{\rm n}})\equiv-\lambda_{0}+\sum_{\mathbf{k}}\left(\frac{1}{2\varepsilon_{\mathbf{k}}}-\frac{1}{2E_{{\rm n}\mathbf{k}}}\right).
\end{equation}
Note that $\Delta_{{\rm o}}$ and $\Delta_{{\rm c}}$ are complex
quantities. Without loss of generality, we set $\Delta_{{\rm o}}$
to be real and positive. From the gap equation (\ref{gapEq}), we
find that $\Delta_{{\rm c}}$ is also real. However, there may exist
two kinds of solutions: an in-phase solution with $\Delta_{{\rm c}}>0$
and an out-of-phase solution with $\Delta_{{\rm c}}<0$. It is easy
to show that for $^{173}$Yb atoms, the out-of-phase solution is responsible
for the BCS-BEC crossover, while the in-phase solution corresponds
to the deep bound state. To show this, we take a look at $\delta=0$
where the two channels become degenerate. In this case, we have $|\Delta_{{\rm o}}|=|\Delta_{{\rm c}}|\equiv\Delta$.
For the out-of-phase solution, the gap equation becomes 
\begin{equation}
\sum_{\mathbf{k}}\left[\frac{1}{2\varepsilon_{\mathbf{k}}}-\frac{1}{2\sqrt{(\varepsilon_{{\bf k}}-\mu)^{2}+\Delta^{2}}}\right]=\frac{M}{4\pi a_{s+}},
\end{equation}
while for the in-phase solution, we obtain 
\begin{equation}
\sum_{\mathbf{k}}\left[\frac{1}{2\varepsilon_{\mathbf{k}}}-\frac{1}{2\sqrt{(\varepsilon_{{\bf k}}-\mu)^{2}+\Delta^{2}}}\right]=\frac{M}{4\pi a_{s-}}.
\end{equation}
Comparing with the two-body result (\ref{Binding0}), we find that
the in-phase solution corresponds to the deep bound state. For this
solution, the chemical potential $\mu$ is large and negative for
all values of the magnetic detuning $\delta$. Therefore, even though
the in-phase solution may be the true ground state of the Hamiltonian,
it is a trivial solution which has nothing to do with the BCS-BEC
crossover associated with the OFR.

The contribution from the Gaussian fluctuations to the thermodynamic
potential can be worked out by completing the path integral over the
fluctuations $\phi$ and $\phi^{*}$. It can be expressed as 
\begin{equation}
\Omega_{\textrm{GF}}=\frac{1}{2\beta}\sum_{Q}\ln\det\left[-\mathbf{\Gamma}^{-1}\left(Q\right)\right],\label{eq:OmegaGF}
\end{equation}
where $Q\equiv(\mathbf{q},i\nu_{l})$ and $i\nu_{l}$ is the bosonic
Matsubara frequencies, and the inverse vertex function (i.e., the
Green function of collective modes) is, 
\begin{equation}
-\mathbf{\Gamma}^{-1}\left(Q\right)=\left[\begin{array}{cccc}
M_{11}^{{\rm o}} & M_{12}^{{\rm o}} & -\lambda_{1} & 0\\
M_{21}^{{\rm o}} & M_{22}^{{\rm o}} & 0 & -\lambda_{1}\\
-\lambda_{1} & 0 & M_{11}^{{\rm c}} & M_{12}^{{\rm c}}\\
0 & -\lambda_{1} & M_{21}^{{\rm c}} & M_{22}^{{\rm c}}
\end{array}\right],
\end{equation}
with the matrix elements at zero temperature (${\rm n}={\rm o,c}$),
\begin{eqnarray}
M_{11,C}^{{\rm n}}\left(Q\right) & = & \sum_{\mathbf{k}}\left(\frac{u_{{\rm n}+}^{2}u_{{\rm n}-}^{2}}{i\nu_{l}-E_{{\rm n}+}-E_{{\rm n}-}}+\frac{1}{2\varepsilon_{\mathbf{k}}}\right)-\lambda_{0},\nonumber \\
M_{11}^{{\rm n}}\left(Q\right) & = & M_{11,C}^{{\rm n}}\left(Q\right)-\sum_{\mathbf{k}}\frac{v_{{\rm n}+}^{2}v_{{\rm n}-}^{2}}{i\nu_{l}+E_{{\rm n}+}+E_{{\rm n}-}},\nonumber \\
M_{12}^{{\rm n}}\left(Q\right) & = & \sum_{\mathbf{k}}\frac{\Delta_{{\rm n}}^{2}}{2}\frac{1/E_{{\rm n}+}+1/E_{{\rm n}-}}{\left(E_{{\rm n}+}+E_{{\rm n}-}\right)^{2}-\left(i\nu_{l}\right)^{2}},
\end{eqnarray}
and $M_{21}^{{\rm n}}(Q)=M_{12}^{{\rm n}}(Q)$, $M_{22}^{{\rm n}}(Q)=M_{11}^{{\rm n}}(-Q)$,
and $M_{22,C}^{{\rm n}}(Q)=M_{11,C}^{{\rm n}}(-Q)$ Here, we use the
short notations $E_{{\rm n}\pm}\equiv E_{{\rm n}\mathbf{k}\pm\mathbf{q}/2}$,
$u_{{\rm n}\pm}^{2}=(1+\xi_{{\rm n}\pm}/E_{{\rm n}\pm})/2$, and $v_{{\rm n}\pm}^{2}=(1-\xi_{{\rm n}\pm}/E_{{\rm n}\pm})/2$.
The summation over the Matsubara frequencies $i\nu_{l}$ in Eq. (\ref{eq:OmegaGF})
is generally divergent. Following the work by Diener \textit{et al.}
\cite{Diener2008}, we cure the divergence by subtracting a \emph{vanishing}
regular term $(k_{B}T/2)\sum_{Q}\ln\det[-\mathbf{\Gamma}_{C}^{-1}(Q)]$,
where $\mathbf{\Gamma}_{C}^{-1}(Q)$ is obtained by replacing $M_{11}^{{\rm n}}(Q)$
with $M_{11,C}^{{\rm n}}(Q)$ and $M_{22}^{{\rm n}}(Q)$ with $M_{22,C}^{{\rm n}}(Q)$,
and by setting $M_{12}^{{\rm n}}(Q)=0$ in $\mathbf{\Gamma}^{-1}(Q)$.
Finally, the convergent result can be expressed as 
\begin{equation}
\Omega_{\textrm{GF}}=\frac{1}{2\beta}\sum_{Q}\ln\left\{ \frac{\det\left[-\mathbf{\Gamma}^{-1}\left(Q\right)\right]}{\det\left[-\mathbf{\Gamma}_{C}^{-1}\left(Q\right)\right]}\right\} .
\end{equation}

In the absence of the inter-channel coupling parameter, i.e., $U_{1}=0$
or $\lambda_{1}=0$, our GPF equations reduce to describe two separate
BEC-BCS crossover Fermi gases in the open and closed channels. In
the unitary limit ($\lambda_{0}=0$), it is known that for each channel
the GPF theory predicts an accurate zero-temperature equation of state
within a few percent relative error \cite{Hu2006,Diener2008}, compared
with the latest experimental measurements \cite{Nascimbene2010,Ku2012}.
At nonzero $\lambda_{1}$, similarly, the GPF theory would be quantitatively
reliable. To solve the EoS at a given detuning $\delta(B)$, we adjust
the chemical potential $\mu$ to satisfy the number equation \cite{Hu2006,Diener2008,He2015-2,Bighin2016}
\begin{equation}
n=-\frac{\partial(\Omega_{\textrm{MF}}+\Omega_{\textrm{GF}})}{\partial\mu},
\end{equation}
and then calculate the pressure $P=-(\Omega_{\textrm{MF}}+\Omega_{\textrm{GF}})$,
compressibility $\kappa=(1/n^{2})(\partial n/\partial\mu)$, and the
speed of sound $c_{s}=\sqrt{n/[m\partial n/\partial\mu]}$. Throughout
this paper, we take $n=5\times10^{13}$ cm$^{-3}$, the typical peak
density for $^{173}$Yb atoms \cite{Pagano2015,Hofer2015}, and $k_{{\rm F}}=(3\pi^{2}n)^{1/3}\simeq1.14\times10^{5}$
cm$^{-1}$, unless otherwise specified. We focus on the \emph{out-of-phase}
solution, which is responsible for the BCS-BEC crossover associated
with the OFR \cite{Zhang2015,Iskin2016}.

\begin{figure}
\begin{centering}
\includegraphics[width=0.48\textwidth]{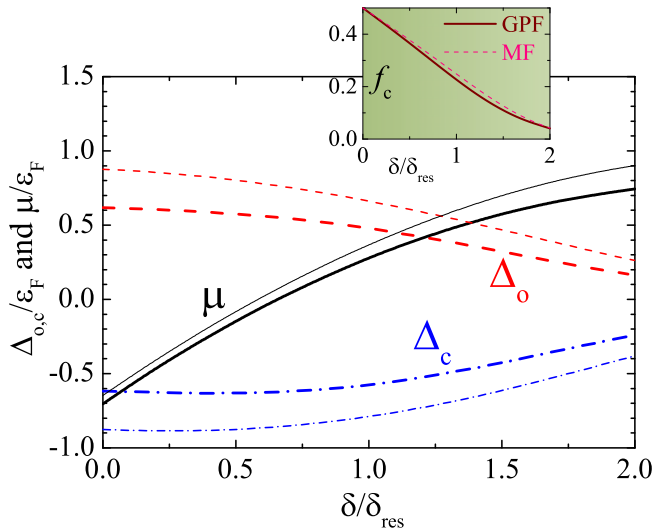} 
\par\end{centering}

\protect\protect\protect\caption{(color online). The chemical potential and two gap parameters as 
functions of $\delta(B)$ at $T=0$. For comparison, the mean-field
predictions are shown by the thin lines. The inset shows the detuning
dependence of the closed-channel fraction $f_{{\rm c}}=n_{{\rm c}}/n$. }

\label{fig3} 
\end{figure}

The solution of $^{173}$Yb atoms from the mean-field theory (MF)
or Gaussian-pair-fluctuation theory (GPF) is shown in Fig. 3, as a
function of the detuning $\delta(B)$ in units of $\delta_{\textrm{res}}=1/(Ma_{s0}^{2})$
\cite{Zhang2015}. The quantitative improvement of our GPF theory
over mean-field is evident and should be observable in future experiments.
Near OFR, the closed-channel fraction is always significant (see the
inset), indicating that the resonantly interacting superfluid may
differ largely from a unitary Fermi gas near a broad MFR \cite{Nascimbene2010,Horikoshi2010,Ku2012}.

%%%%%%%%%%%%%%%%%%%%%%%%%%%%%%%%%%%%%%%%%%%%%%%%%%%%%%%%%%%%%%%%%%%%%%%%%%%%%%%%%%%%%%%%%%%%%%%%

\subsection{Stability of $^{173}$Yb superfluid near OFR}

%%%%%%%%%%%%%%%%%%%%%%%%%%%%%%%%%%%%%%%%%%%%%%%%%%%%%%%%%%%%%%%%%%%%%%%%%%%%%%%%%%%%%%%%%%%%%%%%

\begin{figure}
\begin{centering}
\includegraphics[width=0.42\textwidth]{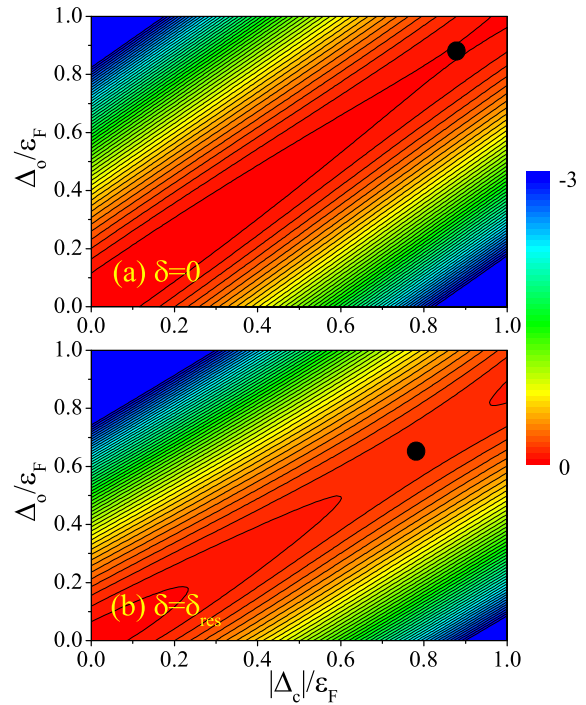} 
\par\end{centering}

\protect\protect\protect\caption{(color online). Contour plots of the mean-field grand potential $\Omega_{{\rm MF}}(\Delta_{{\rm o}},\Delta_{{\rm c}})$
(in units of $n\varepsilon_{\rm F}$) near the out-of-phase solution ($\Delta_{{\rm o}}>0$ and $\Delta_{{\rm c}}<0$)
for $\delta=0$ (a) and $\delta=\delta_{{\rm res}}$ (b). The black
dots indicate the saddle-point positions.}

\label{fig4} 
\end{figure}

The first nontrivial issue we encounter is that the out-of-phase solution
is not a local minimum of the mean-field grand potential $\Omega_{{\rm MF}}(\Delta_{{\rm o}},\Delta_{{\rm c}})$.
In Fig. 4, we show two contour plots of the grand potential (at $\delta=0$
and at $\delta=\delta_{{\rm res}}$). It is clear that the out-of-phase
solution corresponds to a saddle point of the grand potential. The
true ground state corresponds to the deep bound state with energy
$Z_{-}$. In this state, the chemical potential is large and negative,
$\mu\simeq Z_{-}/2$, and hence both two channels are in the deep
BEC state. The BCS-BEC crossover state, which is an exited state,
may suffer from some mechanical instabilities, such as negative compressibility.
We have calculated the compressibility for $^{173}$Yb system. Fortunately,
the compressibility is always positive for the out-of-phase solution.

\begin{figure}
\begin{centering}
\includegraphics[width=0.48\textwidth]{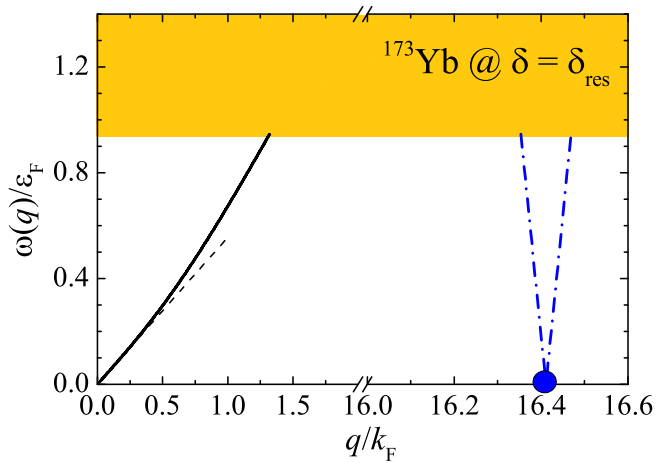} 
\par\end{centering}

\protect\protect\protect\caption{(color online). In-gap density excitation spectrum of a $^{173}$Yb
Fermi gas at the resonance, which touches zero at a large momentum
$q\gg k_{{\rm F}}$ (the big blue dot). The dashed line plots the
linear behavior $c_{s}q$ as $q\rightarrow0$ characteristic of a
sound wave. The colored area indicates the two-particle continuum. }

\label{fig5} 
\end{figure}

Next, we check whether the system suffers from any dynamical instability.
Using the vertex function $\mathbf{\Gamma}(Q)$, it is convenient
to calculate the density excitation spectrum. The dispersions $\omega(q)$
are determined by the pole of $\mathbf{\Gamma}(\mathbf{q},i\nu_{l}\rightarrow\omega+i0^{+})$
after analytic continuation. Below the two-particle continuum, there
are typically two modes corresponding to the in-phase and out-of-phase
fluctuations of the phase of the two order parameters. The in-phase
mode is the well-known gapless Bogoliubov-Anderson-Goldstone phonon
mode, while the out-of-phase mode, predicted by Leggett long 
ago, acquires a finite mass \cite{Leggett1966}. The observation
of a long-lived Leggett mode remains elusive \cite{Blumberg2007,Lin2012,Bittner2015}.

Figure 5 reports the in-gap density excitation spectrum of $^{173}$Yb
atoms. The phonon mode, which behaves like $c_{s}q$ at small momentum,
is clearly seen. However, we are unable to identify a well-defined
gapped Leggett mode. Instead, an anomalous mode is observed at large
momentum $q_{A}\simeq16.4k_{{\rm F}}$. It touches zero and causes
an instability with respect to the density perturbation at the length
scale $l\sim q_{A}^{-1}\simeq5.3$ nm. The existence of such an anomalous
mode is easy to understand. The out-of-phase solution of current interest
is a saddle point solution and hence is intrinsically unstable. We
have checked by varying parameters that the anomalous mode indeed
appears as long as the out-of-phase solution is an excited state [see
Fig. 1(b)]. For the $^{173}$Yb case, fortunately, we do not need to
worry about this dynamical instability, since the nano scale of the
density perturbation is too small to trigger experimentally. Theoretically,
the instability also does not show up in our numerical calculations,
as the pair fluctuation contribution decays exponentially fast with
increasing momentum $q$. Therefore, we conclude that the BCS-BEC
crossover in $^{173}$Yb atoms with OFR is intrinsically metastable
and can be realized in future experiments.

\begin{figure}
\begin{centering}
\includegraphics[width=0.48\textwidth]{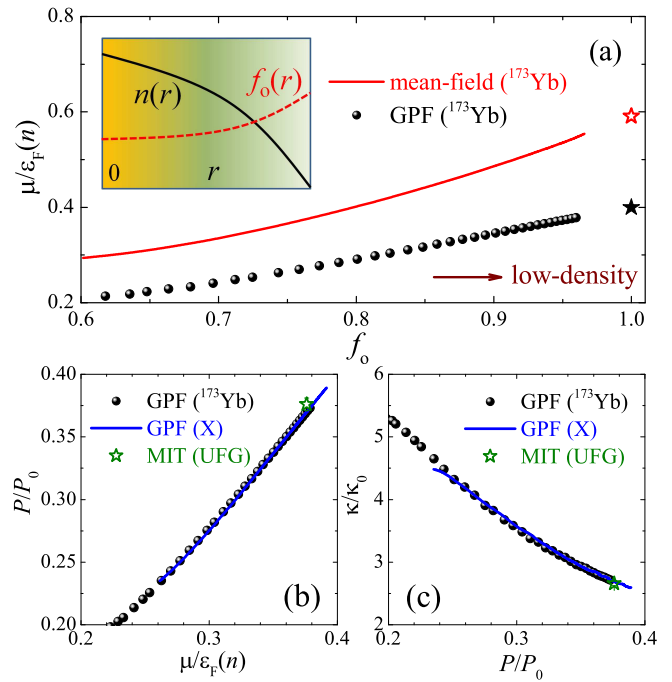}\protect\protect\protect\caption{(color online). (a) The chemical potential of a $^{173}$Yb Fermi
gas at the resonance, as a function of the open-channel fraction with
decreasing density (see the inset for an experimental illustration
in traps). In the low-density limit, where the population of the closed
channel vanishes, the chemical potential approaches the prediction
for broad Feshbach resonances (indicated by stars). Equation
of state, pressure versus chemical potential (b) and compressibility
versus pressure (c), of the resonantly interacting $^{173}$Yb Fermi
gas. The circles are the result for $^{173}$Yb atoms. The blue line
(GPF-X) shows the result for a different set of interaction parameters.
The stars show the MIT result for a unitary $^{6}$Li Fermi gas (UFG)
at broad Feshbach resonances \cite{Ku2012}. The highest density in
our calculations is about $n\sim5\times10^{14}$ cm$^{-3}$.}

\par\end{centering}

\label{fig6} 
\end{figure}

%%%%%%%%%%%%%%%%%%%%%%%%%%%%%%%%%%%%%%%%%%%%%%%%%%%%%%%%%%%%%%%%%%%%%%%%%%%%%%%%%%%%%%%%%%%%%%%%

\subsection{EoS of $^{173}$Yb superfluid at OFR }

%%%%%%%%%%%%%%%%%%%%%%%%%%%%%%%%%%%%%%%%%%%%%%%%%%%%%%%%%%%%%%%%%%%%%%%%%%%%%%%%%%%%%%%%%%%%%%%%

We now explore in greater detail a peculiar feature of the strongly
interacting $^{173}$Yb Fermi superfluid, a peculiar EoS, as a result
of the key component of OFR, the large triplet scattering length $a_{s+}$.
Near the resonance, the grand canonical equation of state, the pressure
$P$ as a function of the chemical potential $\mu$ at $T=0$, can
be expressed as 
\begin{equation}
\frac{P(\mu)}{P_{0}(\mu)}=f_{\mu}\left[\frac{\mu}{\delta_{{\rm res}}};\frac{\delta}{\delta_{\textrm{res}}},\frac{a_{s-}}{a_{s+}},\left\{ x_{i}\right\} \right].
\end{equation}
Here, $P_{0}(\mu)=(2M\mu)^{5/2}/(15\pi^{2}M)$ and $\left\{ x_{i}\right\} $
denotes collectively the other small interaction lengths such as the
effective ranges $r_{s\pm}/a_{s+}$. For $^{173}$Yb atoms, since
the triplet scattering length $a_{s+}$ is large, we may expect that
the dependence on the small parameters $a_{s-}/a_{s+}$ and $x_{i}$
is rather weak. Hence at the resonance, the grand canonical EoS depends
only on the reduced chemical potential $\mu/\delta_{{\rm res}}$,
\begin{equation}
\frac{P(\mu)}{P_{0}(\mu)}\approx f_{\mu}\left(\frac{\mu}{\delta_{{\rm res}}}\right).
\end{equation}

On the other hand, we expect that in the low-density limit $n\rightarrow0$,
or explicitly $\varepsilon_{{\rm F}}/\delta_{{\rm res}}\rightarrow0$,
we recover the universal EoS of the two-component unitary Fermi gas,
which has been realized by using the broad MFR \cite{Nascimbene2010,Horikoshi2010,Ku2012}.
We therefore consider the canonical EoS. The pressure can be expressed
as 
\begin{equation}
\frac{P(n)}{P_{0}(n)}=f_{n}\left[\frac{\mu(n)}{\varepsilon_{{\rm F}}(n)};\frac{\delta}{\delta_{\textrm{res}}},\frac{a_{s-}}{a_{s+}},\left\{ x_{i}\right\} \right].
\end{equation}
For $^{173}$Yb atoms, the dependence on the small parameters $a_{s-}/a_{s+}$
and $x_{i}$ is rather weak. At the OFR we have 
\begin{equation}
\frac{P(n)}{P_{0}(n)}\approx f_{n}\left[\frac{\mu(n)}{\varepsilon_{{\rm F}}(n)}\right],
\end{equation}
where $P_{0}=(2/5)n\varepsilon_{{\rm F}}$. Therefore, the pressure
depends only on a single parameter, the reduced chemical potential
$\mu(n)/\varepsilon_{{\rm F}}(n)$. This peculiar EoS can be easily measured
experimentally. In harmonic traps, all the thermodynamic functions,
in particular, the pressure and compressibility, can be directly determined
from measuring the local density \cite{Ku2012,Ho2010}. Away from
the trap center, with decreasing density, the closed-channel fraction
decreases to zero, due to the enlarged effective detuning, and the
reduced chemical potential $\mu(n)/\varepsilon_{{\rm F}}(n)$ then
increases to reach the universal Bertsch parameter $\xi$ in the broad
MFR limit ($\xi\simeq0.59$ in mean-field theory and $\xi\simeq0.40$
in GPF theory \cite{Hu2006,Diener2008}), as shown in Fig. 6(a). By
varying slightly $a_{s-}/a_{s+}$ and keeping $\delta_{\textrm{res}}$
invariant [i.e., the data labelled GPF-X in Figs. 6(b) and 6(c)], we have
examined theoretically that both $P/P_{0}$ and $\kappa/\kappa_{0}$,
where $\kappa_{0}=3/(2n\varepsilon_{{\rm F}})$, indeed collapse onto
a single curve. We note that, in the dilute limit ($n\rightarrow0$),
we recover the universal EoS of the two-component unitary Fermi gas.
This universal EoS may be understood from the fact that in the dilute
limit, the Zeeman splitting $\delta$ between the two channels becomes
much larger than the Fermi energy $\varepsilon_{{\rm F}}$. In this
case, one can generally show that the closed-channel population becomes
vanishingly small \cite{He2015}. The strong coupling between the
two channels ensures that we recover the universal EoS for the broad
MFR case. However, this universal EoS may hardly be extended to the
high density regime where $n\sim10^{14}$ cm$^{-3}$.

\begin{figure}
\begin{centering}
\includegraphics[width=0.48\textwidth]{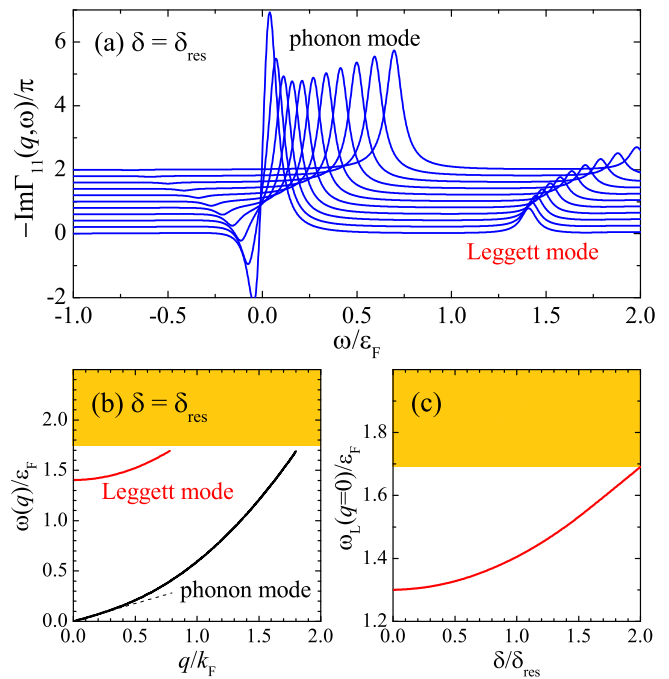} \protect\protect\protect\caption{(color online). (a) In-gap spectral function of Cooper pairs (in arbitrary
units) at the resonance with the scattering lengths $a_{s+}=1900a_{0}$
and $a_{s-}=2a_{s+}$. From bottom to top, the momentum $q$ increases
from $0.1k_{{\rm F}}$ to $1.1k_{{\rm F}}$. The curves are vertically
shifted for better illustration. A finite line width is included to
broaden the $\delta$-peak. (b) The corresponding in-gap density excitation
spectrum. (c) The detuning dependence of the zero-momentum Leggett
mode frequency $\omega_{L}(q=0)$. The colored area shows the two-particle
continuum at $\delta=2\delta_{{\rm res}}$.}

\par\end{centering}

\label{fig7} 
\end{figure}

%%%%%%%%%%%%%%%%%%%%%%%%%%%%%%%%%%%%%%%%%%%%%%%%%%%%%%%%%%%%%%%%%%%%%%%%%%%%%%%%%%%%%%%%%%%%%%%%

\subsection{Leggett mode}

%%%%%%%%%%%%%%%%%%%%%%%%%%%%%%%%%%%%%%%%%%%%%%%%%%%%%%%%%%%%%%%%%%%%%%%%%%%%%%%%%%%%%%%%%%%%%%%%

We turn to consider the condition for the observation of the massive
Leggett mode, by allowing a variable singlet scattering length $a_{s-}$.
It turns out that the out-of-phase solution of the two pairing parameters
becomes the ground state once $1/(k_{{\rm F}}a_{s-})$ is smaller
than a threshold $1/(k_{{\rm F}}a_{s-})_{c}=1/(k_{{\rm F}}a_{s+})$
[see Fig. 1(b)]. It is easy to understand this threshold. Because $a_{s+}=a_{s-}$
at this threshold, the two channels decouple, and hence the out-of-phase
and in-phase solutions become degenerate.

We find that an undamped Leggett mode exists below the two-particle
continuum when $\left|a_{s-}\right|$ is sufficiently large. In this
case, we have two well-behaved condensates that satisfy Leggett's
original picture for the appearance of the massive Leggett mode \cite{Leggett1966}.
Figure 7(a) shows a typical spectral function of the Green's function
of the collective modes for $a_{s-}=2a_{s+}$ and at $\delta=\delta_{\textrm{res}}$,
where the Leggett mode is clearly visible. Its dispersion at small
$q$ can be well approximated by $\omega_{L}^{2}(q)\simeq\omega_{L}^{2}(0)+c_{L}^{2}q^{2}$
[Fig. 7(b)]. With increasing detuning [Fig. 7(c)] or decreasing $1/(k_{{\rm F}}a_{s-})$,
the Leggett mode is pushed upwards, and finally merges into the two-particle
continuum. Experimentally, it is unclear whether we can find a realistic
OFR system with both large singlet and triplet scattering lengths,
which demonstrates the existence of the long-sought Leggett mode.
If such a system can be found, the Leggett mode can be probed by measuring
the dynamic density structure factor via the Bragg spectroscopy \cite{Lingham2014}.

%%%%%%%%%%%%%%%%%%%%%%%%%%%%%%%%%%%%%%%%%%%%%%%%%%%%%%%%%%%%%%%%%%%%%%%%%%%%%%%%%%%%%%%%%%%%%%%%

\section{Summary}

\label{s5} %%%%%%%%%%%%%%%%%%%%%%%%%%%%%%%%%%%%%%%%%%%%%%%%%%%%%%%%%%%%%%%%%%%%%%%%%%%%%%%%%%%%%%%%%%%%%%%%

In summary, we calculated the OFR with realistic Lenard-Jones potentials
and presented a low-energy effective theory for OFR which is useful
for field theoretical study of the many-body system. We presented
a strong-coupling pair fluctuation theory for the BCS-BEC crossover
in $^{173}$Yb atoms across its OFR. The stability of the BCS-BEC
crossover, the equation of state at the OFR, and the collective modes
(in particular the massive Leggett mode) are investigated by using
the pair fluctuation theory. Since the BCS-BEC crossover in $^{173}$Yb
atoms corresponds to an excited state, there exists a dynamical instability
with respect to an inhomogeneous density perturbation. Fortunately,
due to the small singlet scattering length, this instability occurs
at very large momentum and hence is hard to trigger under current
experimental conditions. Hence the BCS-BEC crossover in $^{173}$Yb
atoms with OFR is intrinsically metastable and can be realized in
future experiments. The small singlet scattering length in $^{173}$Yb
atoms also leads to a peculiar EoS, which is peculiar for a Feshbach
resonance with sizable closed-channel fraction. The massive Leggett
mode in the superfluid state of $^{173}$Yb atoms is severely damped.
We find that an undamped Leggett mode exists only for the case with
both large singlet and triplet scattering lengths.

Our quantitative predictions could be experimentally examined in the
near future in cold-atom laboratories \cite{Pagano2015,Hofer2015}.
They also might be relevant to other two-band fermionic superfluids and superconductors
in diverse fields of physics, such as MgB$_{2}$ and LaFeAsO$_{0.89}$F$_{0.11}$
in solid-state physics \cite{Xi2008,Hunte2008}.

\begin{acknowledgments}
We would like to thank Hui Zhai for useful discussions. This work
is supported by the Thousand Young Talent Program in China (L.H.), the
ARC Discovery Projects: DP140100637 and FT140100003 (X.J.L.) and FT130100815
and DP140103231 (H.H.), and the National Natural Science Foundation of China,
Grant No.11474315 (S.G.P.). 
\end{acknowledgments}

\end{document}